\documentclass{pnastwo}

\usepackage{graphicx}

\usepackage{amssymb,amsfonts,amsmath,citesort}

\begin{document}




\track{ACCEPTED FOR PUBLICATION IN PNAS}



\footlineauthor{Hsu, Largo, Sciortino, \& Starr}

\url{www.pnas.org}
\issuenumber{no. xx}
\volume{vol. 105}
\issuedate{Appearing online Week of Aug.~25, 2008}

\title{Hierarchies of networked phases induced by multiple liquid-liquid critical points}




\author{
Chia Wei Hsu\affil{1}{Department of Physics, Wesleyan University, Middletown, Connecticut 06459, USA},
Julio Largo\affil{2}{Dipartimento di Fisica and CNR-INFM-Soft: 
Complex Dynamics in Strucured Systems, Universit\`a di Roma La Sapienza, 
Piazzale Aldo Moro 2, I-00185 Rome, Italy}\affil{3}{Departamento de Fisica Aplicada, Universidad de Cantabria, Avda. Los Castros s/n Santander, 39005, Spain.},
Francesco Sciortino\affil{2}{},
\and
Francis W. Starr\affil{1}{}}

\contributor{Accepted for publication in the Proceedings of the National Academy of Sciences 
of the United States of America}

\maketitle

\begin{article}

\begin{abstract} 
Functionalization of nanoparticles or colloids is increasingly being used to develop customizable ``atoms''.  Functionalization by attaching single strands of DNA allows for direct control of the binding between nanoparticles, since hybridization of double strands will only occur if the base sequences of single strands are complementary.
Nanoparticles and colloids functionalized by four single strands of DNA can be thought of as designed analogs to tetrahedral network forming atoms and molecules, with a difference that the attached DNA strands allow for control of the length scale of bonding relative to the core size.  We explore the behavior of an experimentally realized model for a nanoparticles functionalized by four single strands of DNA (a tetramer), and show that this single-component model exhibits a rich phase  diagram with at least three critical points and four thermodynamically distinct amorphous phases.  We demonstrate that the additional critical points are part of the Ising universality class, like the ordinary liquid-gas critical point. The dense phases consist of a hierarchy of interpenetrating networks, reminiscent of a woven cloth.   Thus bonding specificity of DNA  provides an effective route to generate new nano-networked materials with polyamorphic behavior.   The concept of network interpenetration helps to explain the generation of multiple liquid phases in single-component systems, suggested to occur in some atomic and molecular network forming fluids, including water and silica.
\end{abstract}




\dropcap{T}echnological developments are creating a wide variety of  new building blocks  which play the role of  ``functionalized atoms''  for designed materials~\cite{glotzer,frenkel}.  
Borrowing ideas from biological specificity~\cite{condon06,niemeyer,berti,seeman2}, it is now possible to control bonding between engineered  building blocks in a selective way, moving in the direction of  an effective synthetic bottom-up strategy for self-assembly.  
 The complementary binding of base pairs,  combined with the ability to  directly control the base sequence,  makes DNA an ideal candidate for the  development of network-based,  nanostructured
  materials~\cite{condon06,seeman2,mirkin,chaikin}.
A variety of experiments~\cite{mirkin,chaikin,gang08,mirkin08,seeman3,mclaughlin,baglioni,kiang05,crocker05} 
have demonstrated the possibility to realize nanostructured, DNA-based materials, including ordered crystal structures~\cite{mirkin08,gang08}.  

By grafting short ({\it i.e.}\ 
smaller than the persistence length) single strands (ss) of DNA
to core particles, the core can act as a ``node'' of a complex network
formed when single strands combine into double-stranded (ds) DNA.  Longer linking ``arms'' may be achieved by using a double-stranded spacer near the core nanoparticle~\cite{mirkin08,gang08}.
 With an intelligent choice of the core, it is possible to control the nearest neighbor coordination of the  functionalized nanoparticle, providing a direct route to study materials where  the bonding coordination is much less than that of spherically symmetric molecules~\cite{zac05}. 
In such limited coordination systems, the general features
of network formation are expected to be observed. 

By functionalizing the central core with four DNA strands, we expect to
generate an engineered analog of tetrahedral network forming atoms or molecules, in which the bonding contribution to the interparticle interaction energy
can be tuned by modulating the length $L$ of the DNA strand, as well as the chemical properties of the solvent.  Tetrahedrally coordinated fluids, such as water~\cite{pses92,ms98,ms98review,debenedetti-review,brovchenko}, silicon~\cite{sa03}, and silica~\cite{saika} appear to have unusual phase diagrams that include two critical points: (i) the usual gas-liquid critical point, and (ii) a lower temperature critical point terminating the coexistence of a low-density and high-density liquid.  This phenomenon is also referred to as ``polyamorphism''~\cite{polyamorphism}. Our model DNA ``{molecules}'' allow us to explore if such unusual polyamorphic behavior is shared by designed systems, and also provide insights that might help understand complex phase diagrams in traditional materials.  
{Our findings indicate that interpenetration of repeating, amorphous networks can give rise to multiple fluid phases, providing a disordered analog of the 
natural example of interpenetration which occurs in ices VI, VII, and VIII~\cite{fletcher}, and which helps to understand how interpenetration in liquid water is feasible at high pressures.  }

Our results are based on a series of Monte Carlo simulations of a
 simple, coarse-grained molecular  model
for tetrahedral DNA dendrimers -- that is, a core nanoparticle binding four ss-DNA with tetrahedral symmetry. Such tetramers have also been experimentally synthesized~\cite{mclaughlin}.  The  model retains two 
important basic features: (i)
base-pair selectivity and (ii) the limitation that 
each ss-DNA can  bind to only one other ss-DNA to form a ds-DNA. The model omits chemical
details and explicit solvent effects which would make calculations of the phase behavior infeasible.  Each molecular unit of the model is composed of a tetrahedral hub tethering four identical DNA-like strands composed of $L$ connected monomers.
The detailed model (introduced in refs.~\cite{ss,lss}) is shown for the case $L=8$ in Fig.~1 A and B.
 
\begin{figure}
\noindent
\includegraphics[width=3.5in]{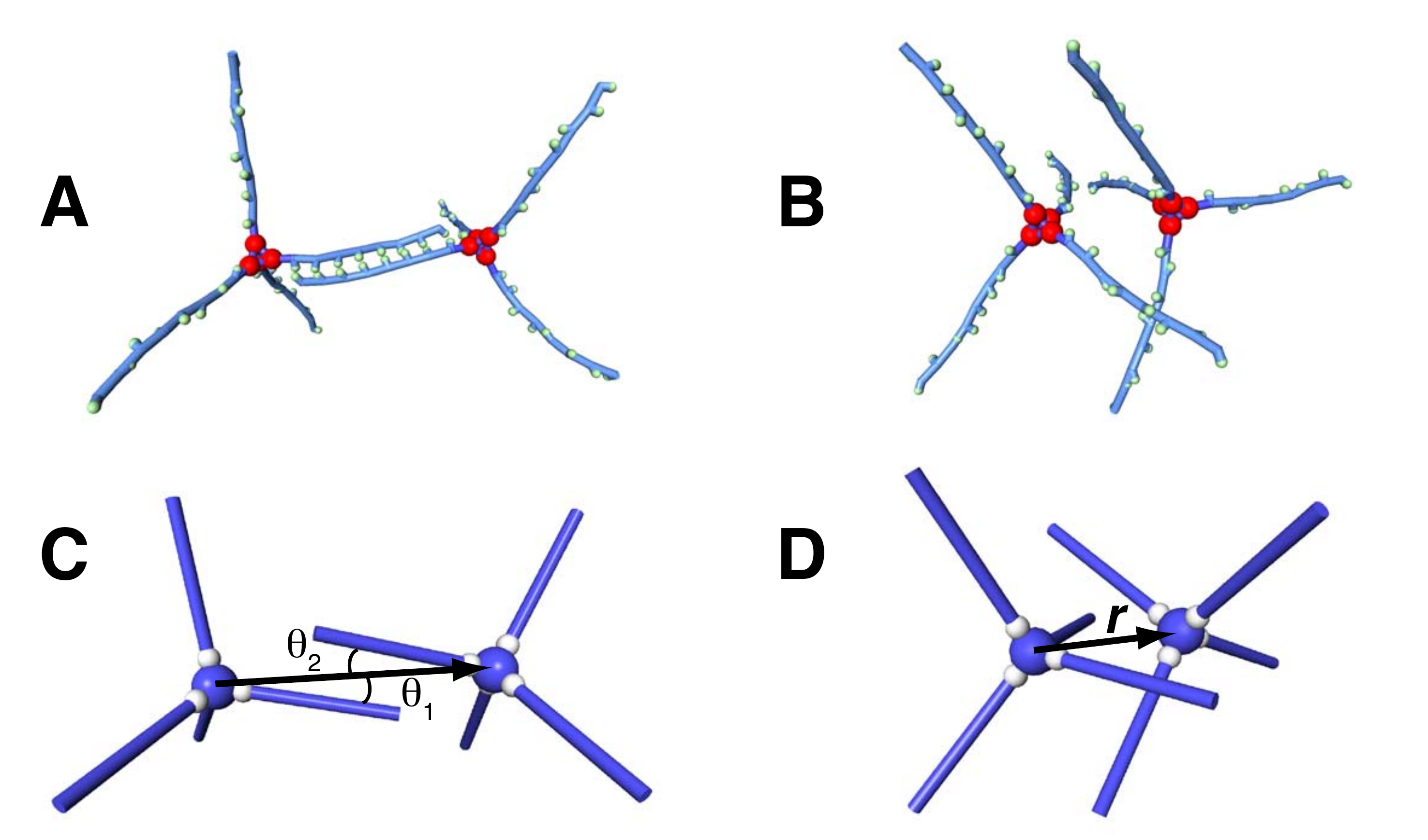}
\caption{Visualization of the model before and after coarse graining.  A single dendrimer is formed by  joining four single strands to a central hub (the four red  spheres).   The small green spheres are the attractive ``sticky'' spots that  carry information about base type. The top panels show (A) a bonded pair of dendrimers and (B) a nearby pair that is unbonded.   In the coarse-grained model (bottom panels), the 4 central sites are replaced by a single core (indicated by a blue sphere), and the arms (indicated by a blue cylinder),  each representing a single strand of DNA, the arms are parameterized by a distance and angle dependent bonding potential. The schematics show both (C) bound and (D) unbound cases of two nearby dendrimers.  The potential between dendrimers depends on the core-core separation $r$ and the orientations $\theta_1$ and $\theta_2$ of the binding arms relative to $r$. }
\label{fig:fig1}
\end{figure}

Here we focus mainly on a simplified model~\cite{lst} that is directly derived from the more complex model.  We represent each dendrimer by a central core with rigid arms -- representing the ss-DNA -- that point to the vertices of tetrahedron (Fig.~1 C and D).  The coarse-grained dendrimer-dendrimer interaction potential is a function only of (i) the separation $r$ of the cores of the nanoparticles, (ii) the  relative orientations $\theta_1$ and $\theta_2$ of the binding strands, and (iii) the temperature $T$ dependence of the probability $p_B(T)$ that single strands bond to form double strands~\cite{lst}.   Fig.~2 shows both the separation dependence of the potential along the bonding direction ($\theta_1=\theta_2=0$) and the angular dependence at the bonding distance.  The details of the potential and the parameterization from the more detailed model are described in the Methods section. 
In contrast with atomic or molecular network forming systems, the core-core excluded volume interactions of our model are relevant only at separations much smaller than the bonding distance, allowing for significant interpenetration effects.  The model was originally parameterized for DNA sequences of length $L=8$~\cite{lst}; here we generalize the model to shorter  ($L=4$) and longer ($L=16$) strands. The sequence of the DNA bases in each arm is chosen so that the complementary sequence is identical, but in reverse order.  As a result head-to-tail bonding  between the arms of different dendrimers occurs, and when many dendrimers bind, they form an open tetrahedral network.

\begin{figure}[t]
\noindent
\includegraphics[width=3.5in]{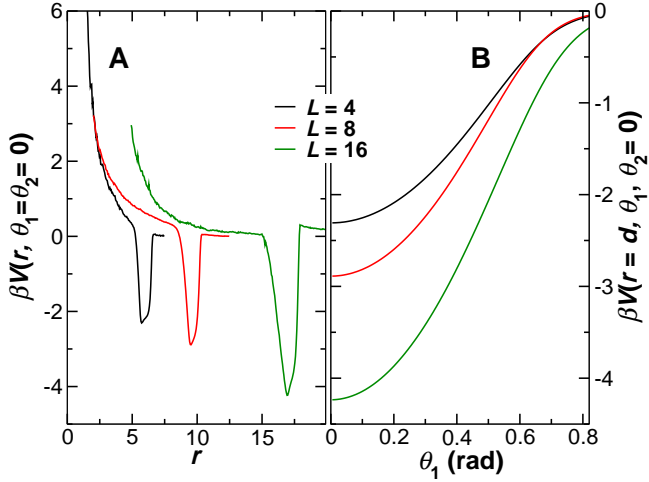}
\caption{Dendrimer-dendrimer pair potential for the model studied.    (A) The potential $\beta V(r,\theta_1 , \theta_2=0)= \beta (V_{NB}(r) + V_{B}(r,\theta_1 , \theta_2=0))$ along the ideal bonding direction, where $\beta = 1/k_B T$.   For simplicity of the figure, we take $p_B = 1$, so that the full effect of the bonding potential is apparent. (B) The angular variation of the potential at the ideal bonding distance.  The attraction drops rapidly as the $\theta_1$ departs from zero.  Note that since labels are arbitrary, $V(r=d,\theta_1=0 , \theta_2)$ vs. $\theta_2$ is identical to $V(r=d,\theta_1 , \theta_2=0)$ vs $\theta_1$.}
\label{fig:fig2}
\end{figure}

\section{Results}

To evaluate the phase behavior, we first determine the location of 
any critical points in the phase diagram.  In the grand canonical ensemble, we can accurately calculate the position of a critical point  by locating the state points (defined by chemical potential $\mu$ and $T$) where the density fluctuations
exhibit a specific bimodal distribution. 
We define the density  $\rho \equiv  N/V d^3$,  where $d \approx L$  is the most probable separation of nearest neighbor dendrimer cores when they 
  are bonded, $N$ is the number of dendrimers and $V$ is the volume.  Including $d$ in the definition of $\rho$ simplifies comparison of systems with different $L$.
Surprisingly, as shown in Fig.~3, we discover  {\it three} different critical points for all investigated $L$, namely: the expected critical point terminating the transition between the unassociated dendrimers and a networked fluid state (the analog of the gas-liquid critical point in fluids) as well as two additional liquid-liquid critical points.    To demonstrate that all three critical points belong to the Ising universality class, we show that
the calculated order parameter distribution $P(M)$, where $M = \rho -su$,
coincides with the known distribution of the magnetization in the Ising model at the 
critical point~\cite{ising-CP}.   Here, $u$ is the potential energy density and $s$ is referred to as the field-mixing parameter.  
The possibility of multiple critical points in networked liquids is well documented~\cite{pses92,ms98,ms98review,debenedetti-review,brovchenko,sa03,saika,phosphorus,bs03}, but it has never before been possible to quantitatively observe the critical behavior and identify the universality class of any additional transitions.  Additionally, the data in fig.~3 unambiguously show that there may be a hierarchy of many such critical points.

\begin{figure}[t]
\noindent
\includegraphics[width=3.5in]{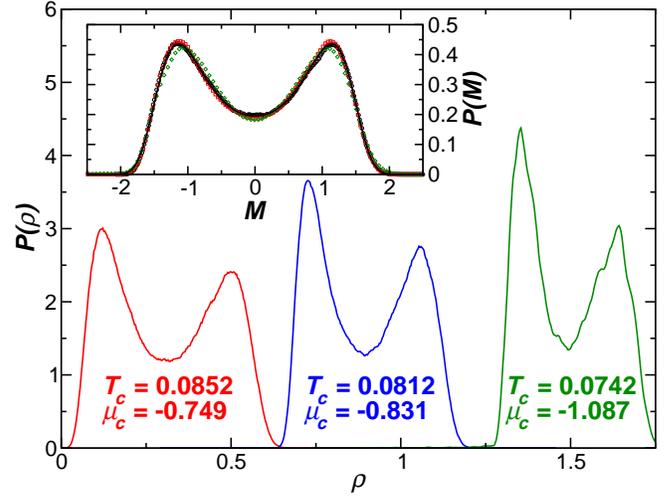}
\caption{The distribution of the   density $P(\rho)$ for $L=4$ at the locations ($T_c$, $\mu_c$)
  of the three distinct critical points terminating first-order fluid-fluid transitions.   
  In the inset,   we identify the universality class  by calculating the distribution    of the order parameter $P(M)$, which    
coincides with the known distribution for the Ising model (solid black line).   
Note that $M$ is shifted by its mean value and scaled
  so that it has unit variance.}
\label{fig:fig3}
\end{figure}

To  complete the picture of the phase behavior, we identify the coexisting phases and evaluate the phase boundaries (Fig.~4 A-C).  For each $L$, three adjacent phase coexistence regions emerge, with roughly same
width in $\rho$, suggesting the presence of well-defined preferential $\rho$ values for the stable dense phases.  The $\rho$ dependence for all $L$ can be better understood by scaling $\rho$ by the corresponding ideal diamond lattice density $\rho_d = 3\sqrt{3}/8 \approx 0.650$ (in units where the nearest neighbor lattice sites are separated by unit length)~\cite{kittel}. 
In this representation, the densities of the dense stable phases are approximatively 
integer multiples of $\rho_d$, suggesting that these phases consist of interpenetrating four-coordinated networks.
For sufficiently large $L$, where the core size is very small compared to the bonding distance, we expect the formation of additional interpenetrating networks, and of their associated critical points, beyond those we have already observed.  So far our studies have not identified such a scenario, due to the excessive computations required to equilibrate at large densities and small temperatures.
 Unlike density, scaling of the temperature to a universal form is more complicated. Indeed,  the
differences in the $L$ dependence of the bonding   entropy  causes the relative value of the critical
temperature $T_c$ to differ.  A graphical representation of the dense stable phases is shown in
Fig.~4 D-F.

\begin{figure}[t]
\noindent
\includegraphics[width=3.5in]{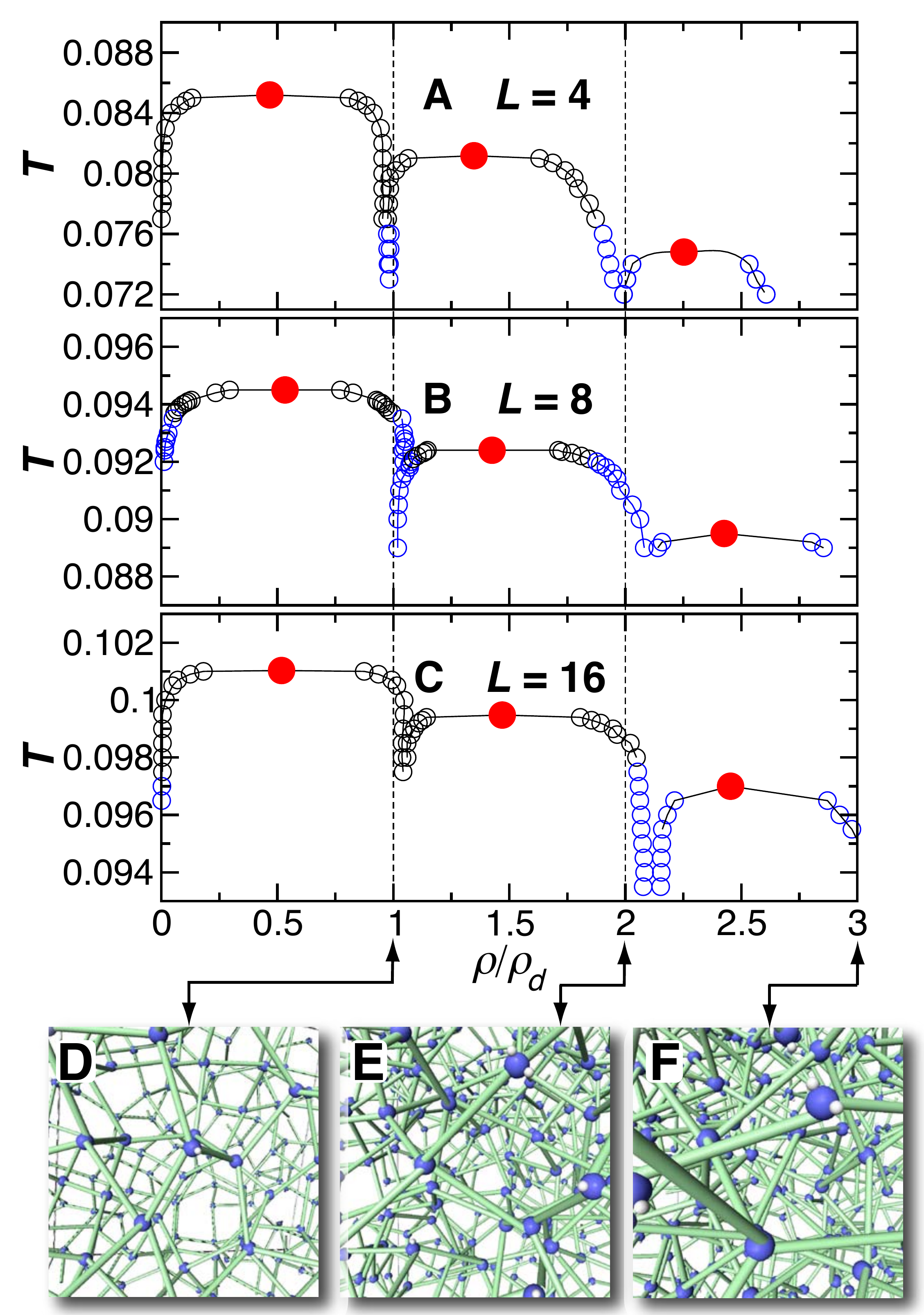}  
\caption {Phase diagrams and schematics of fluid structure. We plot the phase diagrams for arms of length (A) $L=4$, (B)  $L=8$ and (C) $L=16$ bases.  The red circles indicate the critical point locations.    The black
  symbols indicate the phase boundary determined by a histogram
  reweighting procedure~\cite{wilding}, and the blue symbols are
  determined from grand canonical ensemble simulations.  We illustrate the structure  of the dense fluids in panels (D-F) which
  correspond the the low, medium, and high density networked states, as
  indicated by the arrows.    Green cylinders are drawn between the
  centers of bonded dendrimers.  The arms of the
  dendrimers are indicated only by a small white sphere to
  provide the molecular orientation.}
\label{fig:fig4}
\end{figure}

To quantify how interpenetrating networks are manifested in  typical measured structural properties, and to provide evidence for the absence of crystal order, we evaluate the radial distribution $g(r)$ 
(where $r$ is the separation between the centers of dendrimers)
and its Fourier transform, the structure factor $S(q)$ (Fig.~5 A and B). The tetrahedral order of the lowest density networked phase is evidenced by the fact that the ratio of the position of the first and second neighbor peaks of $g(r)$ is approximately 
$ 4/\sqrt{6} \approx 1.63$, expected for ideal tetrahedral coordination.  The second and third interpenetrating liquids also have the same ratio between first and second neighbor peak locations, but there is increased amplitude of $g(r)$ for $r/d < 1$. Such an increase is consistent with  the presence of dendrimers filling the holes of the complementary networks.
The presence of  neighbors closer than the bonding distance is made 
possible by the absence of significant hard-core interactions.  Similar tetrahedral signatures are indicated by $S(q)$.  For the single-networked fluid, the correlation among the holes of the network gives rise to the ``prepeak'' in $S(q)$ at $q/q_0\approx \sqrt{6}/4$, where $q_0$ is the location of the main peak of $S(q)$.  As these holes are filled by subsequent interpenetrating networks, the prepeak weakens, making it difficult to recognize from $S(q)$ the underlying tetrahedral structure.

To demonstrate that even the network of the densest phase is locally tetrahedral and interpenetrating, we look at the contribution to the $g(r)$ 
restricted to dendrimers that are separated by  a specific number of bonds, refereed to as
chemical distance $D$ (Fig.~5 C). More specifically, $g(r)$ for  $D=1$ only shows correlations with the nearest connected neighbors, while $D=2$ includes both first and second connected neighbors, etc. 
In this way, we consider only neighboring particles that both belong to the same bond network as the selected central particle and are separated by no more than $D$ bonds.
 The ordinary $g(r)$ is recovered in the limit $D\rightarrow\infty$.  Fig.~5 C shows that for $D\le 3$, the system has nearly perfect tetrahedral order and there is almost no interpenetration.  For $D>3$, connections between networks  destroy the long range order and dendrimers can loop back to very small physical distance $r$, giving rise to the increase  in the amplitude at $r/d<1$.  
We visualize the interpenetrating structure of the three networks in the densest liquid by rendering dendrimers and the bonds between dendrimers of distinct networks with distinct colors (Fig.~5 D-F).

 \begin{figure*}[t]
\noindent
\includegraphics[width=7in]{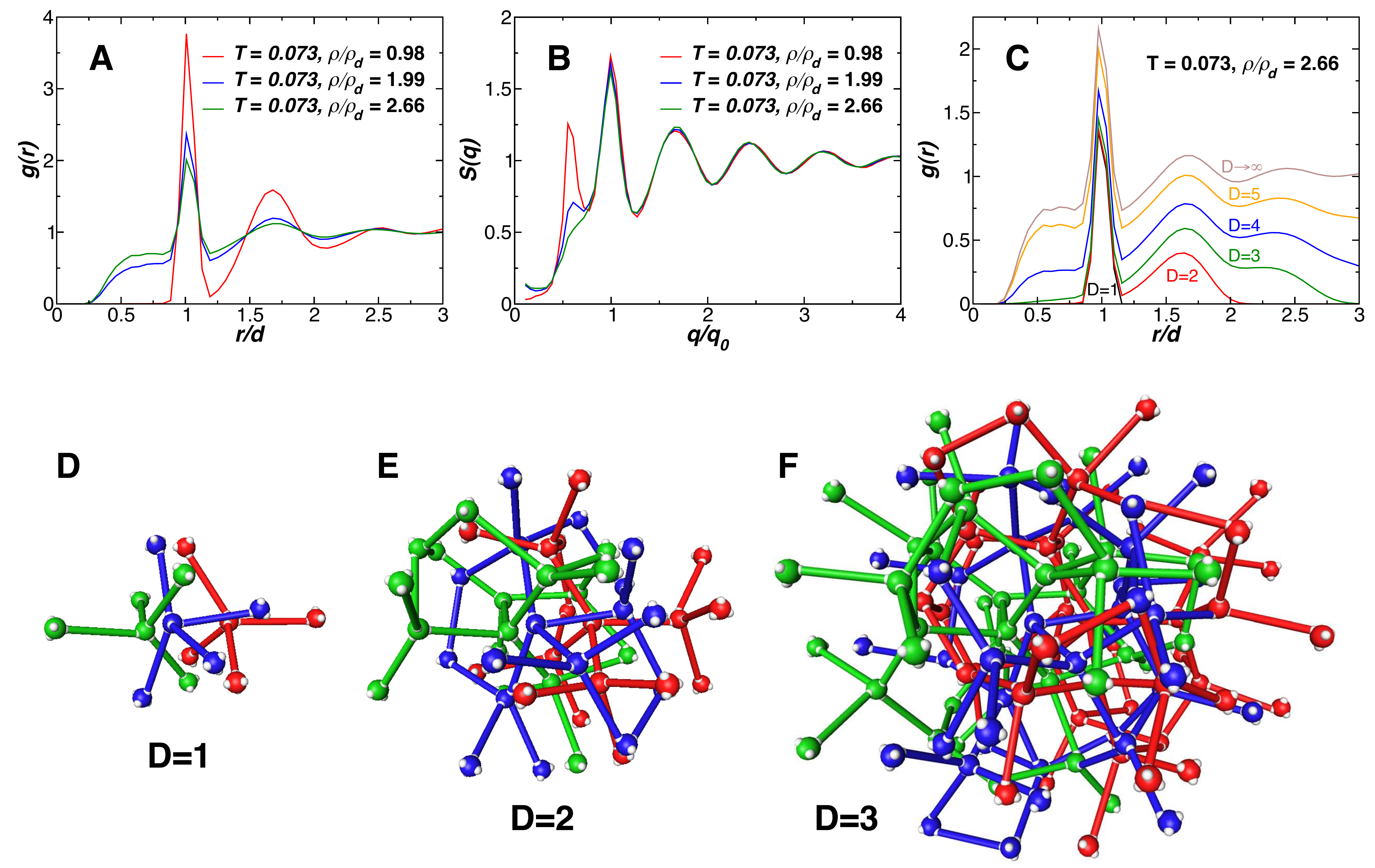}
\caption{Demonstration of the existence of interpenetrating networks and quantification of their influence on traditional structural measures.    (A) The radial distribution $g(r)$ as a function of scaled distance $r/d$ for three distinct liquids phases.  The increased amplitude for $r/d<1$ is a consequence of the interpenetration.  (B)  The structure factor $S(q)$ for the same three liquids.  (C) $g(r)$ restricted to bonded neighbors up to a specific chemical distance $D$,  providing evidence for the tetrahedrality of the interpenetrating networks.  (D-F)  Visualization of the state point used for calculations of {\bf c} to show the three interpenetrating networks.   Specifically,  for an arbitrary selected dendrimer and for the two closest unbonded neighbors of that dendrimer, we draw their bonded neighbors up to  $D=3$.}  
\label{fig:fig5}
\end{figure*}

Lastly, we consider the possible concern that additional phase transitions may be only an artifact of the simplified model.  Since comprehensive phase diagram calculations for the original model of fig.~1 (A,B) are infeasible, we instead check for phase separation in both the original and effective potentials by simulating at fixed total $\rho$ in the phase coexistence region.  Fig.~6 shows $S(q)$ after a long equilibration to allow phase separation to complete for both the original and effective potentials in the liquid-gas coexistence region, as well as the first liquid-liquid coexistence region.  The growth of $S(q)$ for small $q$ is an indication of phase separation.  The growth is much more pronounced for the liquid-gas coexistence because the ratio of densities of the coexisting phases is much larger than for the liquid-liquid coexistence.  The behavior of $S(q)$ for the real and effective potentials is nearly quantitatively identical, indicating that the same phase separation process occurs in the original potential.

\section{Summary}

The emergence of multiple critical points in this system results from the possibility of creating multiple interpenetrating bonded structures,  explaining why  such phenomenon is not observed  in simple atomic and molecular systems for which hard-core interactions prevent significant interpenetration.  However, interpenetration is in principle possible in some networked liquids  where directional bonding  favors locally ``empty'' bonded states, like water.  In this respect, our work supports the possibility of multiple liquid-liquid critical points in water and other networked liquids as a result of interpenetration of distorted bonded networks. 
{ Some care must be taken in applying interpenetration concepts to water and other molecular systems.  Specifically, the hydrogen bond distance in water is not substantially larger than the core (oxygen-oxygen) repulsion.   Despite this repulsion, water is able to arrange, at least in crystal forms VI, VII, and VIII, as two interpenetrating fully bonded lattices~\cite{fletcher}.  Unlike our DNA functionalized particles, the absence of bond-selectivity in water molecules means that molecules can share their bonding sites to form simultaneously pairs of  bonds~\cite{sgs91}.   These shared bonds are absent for DNA functionalized particles, where the lock-and-key mechanism imposes a one-to-one correspondence between bonding sites and number of bonded neighbors.   Such shared bonds in water give rise to significant distortions of the network, which is clearly indicated from structural measurements~\cite{sbs}.  As a result, in the case of water, the density of the high density interpenetrating phase should be significantly smaller than twice that of a single networked phase.  Indeed, the difference between the low- and high-density states is experimentally known to be around 25\%~\cite{lg06}.  Another approach that has helped to understand multiple phases in these dense systems is the use of spherically symmetric models with two length scales~\cite{ms98review,franzese01,xu05,xu06}}.  Here we have shown that a single bonding length scale that allows for interpenetration is sufficient to understand multiple amorphous phases, providing a complementary approach.  Moreover, our results demonstrate that novel nanostructured materials can serve as model systems to help understand the behavior of conventional materials.


\begin{figure}[h]
\noindent
\includegraphics[width=3.4in]{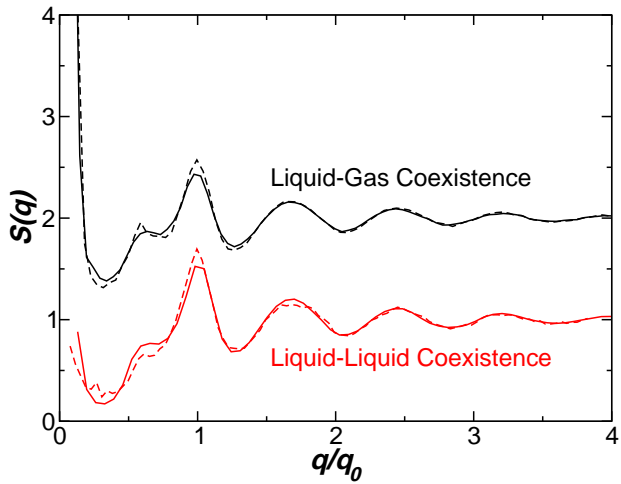}
\caption{Comparison of the structure for the original and effective potentials in the phase coexistence region for $L=4$.  We calculate $S(q)$ only after a long equilibration to allow the phase separation process to near completion. The solid line is the effective potential, and the dotted line is the original potential of refs.~\cite{ss,lss}. The system density is fixed so that the system phase separates into coexisting phases.  For the liquid-gas coexistence, we study $T=0.083$ and $\rho/\rho_d = 0.58$; for the liquid-liquid coexistence, we study $T=0.079$ and $\rho/\rho_d = 1.42$, which can be readily identified in fig.~4 A. }  
\label{fig:fig6}
\end{figure}

In summary, the chemical synthesis of new functionalized nanoparticles offers the possibility to control the bonding orientation and to generate bonding distances which are significantly larger than the core size,  a key element to favor the formation of multiple interpenetrating networks at high density. DNA functionalized cores provide a very effective way to realize such nanoparticle building blocks and to create structured materials with properties that do not ordinarily appear in nature.  
Thus, our work suggests that it is possible to expand the tool-box of building blocks so as to open the possibility of a hierarchy of networked phases, which could potentially be amorphous or ordered.  In the present case, the interpenetrating networks are identical.  However, by appropriate mixing of particles functionalized with distinct sequences of DNA, it should be possible to generate interpenetration in which each network has distinct properties.   Such a material would provide three-dimensional nanoscopic canvass, woven by differently ``colored'' wires, the properties of which can be controlled directly by base sequence selection.

\section{Methods}

We first briefly describe the model and procedure for obtaining the parameters of the effective potential we use.  
 The potential we study is a directly derived from a more detailed 
model~\cite{ss,lss} for four-armed DNA dendrimers. 
Obtaining potential parameters for our coarse-grained representation requires an initial study of high and low $T$ states in the more complex model.  The more complex model includes explicit bases, and we choose a
palindromic sequence of bases to ensure head-to-tail bonding that gives rise to large networks.  The bases comprise of four types of particles labeled  A, C, G, and T. Bases of type A bond only with type T, and type C bonds only with type G. Starting from the tetrahedral core, the specific ordering of the bases used to determine the parameters of the effective potential are: (i) $L=4$, A-C-G-T; (ii) $L=8$, A-C-G-T-A-C-G-T; (iii) $L=16$, A-A-G-C-C-A-G-T-A-C-T-G-G-C-T-T.

The coarse graining procedure has been explained in detail in Ref.~\cite{lst}.  We express the effective pair potential as sum of the
potential for the bonding state $V_{B}$ (weighted by the probability $p_B$ that two strands are bonded at temperature $T$)
and the potential describing the non-bonding interaction $V_{NB}$, {\it i.e.}
\begin{equation}
V(r,\theta_1,\theta_2,T)=V_{NB}(r)+p_B(T)\, V_{B}(r,\theta_1,\theta_2).
\end{equation}
The potential $V_{NB}$ is  assumed to be spherically symmetric and it is very soft since it models  primarily  the repulsion of the small core. The potential $V_B$ retains orientation dependence,
since bonding only occurs when strands are properly aligned.   To mimic  bond selectivity  and to reproduce the lock and key binding characteristic of biological binary associations like DNA, each arm may form no more than one bond.
Therefore, if the arms labeled by $\theta_1$ or $\theta_2$ 
are already engaged in a bond with some other dendrimer in the system, $p_B(T)=0$.

To numerically evaluate the functional form of $V_{NB}$, $V_{B}$ and $p_B(T)$  we simulate a system composed of two dendrimers
in a box for several  $T$ in the
region where ss-DNA pairs. 
The numerical results for the probability $p_B(T)$ can be modeled as a  two-state  expression  \begin{equation}
p_b(T)=\left(1+\exp\left({-\frac{\Delta U - T \Delta S}{k_B T}}\right)\right)^{-1},
\end{equation}
where $\Delta U$ and $\Delta S$ measure the change in energy and entropy associated to the formation of a double strand and $k_B$ is the Boltzmann constant (in our reduced units, $k_B=1$).   
The angular dependence can be model as a Gaussian function.
The ability of the effective potential to reproduce structural and thermodynamic properties 
of the explicit model has been verified in ref.~\cite{lst} for the case $L=8$. 
We have confirmed that similar agreement in structural quantities is observed for the case $L=4$ and  $L=16$,
even in the region where multiple networks are observed.


To evaluate the phase diagram, we carry out a series of Monte Carlo simulations in the grand canonical ensemble (fixed $\mu$, volume V, and T)
and study the distribution of the density fluctuations. To properly locate the  critical points, we use histogram re-weighting techniques to ensure the order parameter distribution $P(M)$ matches that of the Ising universality class.    The phase boundaries near the critical point are determined by a series of multi-canonical simulations, as described by ref.~\cite{wilding}.

\bigskip

\noindent {\bf Acknowledgments}~~ We thank J.F.~Douglas and S.~Sastry for helpful discussions. We acknowledge the NSF for support under Grant No. DMR-0427239, and we thank Wesleyan University for computer time supported by the NSF under grant number CNS-0619508.
We also acknowledge support from MIUR, MCRTN-CT-2003-504712, and MERG-CT-2007-046453.


\end{article}










\begin{thebibliography}{99}

\bibitem{glotzer} 
Glotzer SC, Solomon MC (2007)
\newblock Dimensions in anisotropy space: rationalizing building block
  complexity for assembly.
\newblock {\em Nature Materials} 6: 557-562.

\bibitem{frenkel}
Frenkel D (2006) Colloidal crystals: Plenty of room at the top. {\em Nature Materials} 5: 85-86.



\bibitem{berti} 
Berti D  (2006) Self assembly of biologically inspired amphiphiles
{\em Current Opinions in colloid and interface science} 11: 74-78.

\bibitem{niemeyer}
Niemeyer CM (2000) Self-assembled nanostructures based on DNA: towards the development of nanobiotechnology
{\em Current Opinion in Chemical Biology} 4: 609-618.

\bibitem{seeman2}
Seeman NC (2003) DNA in a material world. {\it Nature} 421: 427-431.

\bibitem{condon06}
Condon A (2006) Designed DNA molecules: principles and applications of molecular nanotechnology. {\em Nature Reviews Genetics} {7}: 565-575.


\bibitem{chaikin} Valignat MP,  Theodoly O,  Croker JC,  Russel WB, Chaikin PM
(2005) \newblock Reversible self-assembly and direct assembly of DNA-linked micrometer-sized colloids
{\em Proc. Nac. Academy Science} {102}: 4225-4229.

\bibitem{mirkin} Mirkin CA, Letsinger RL, Mucic RC, Storhoff JJ (1996)
A DNA-based method for rationally assembling nanoparticles into macroscopic materials 
{\em Nature} {382}:  607-609.


\bibitem{mirkin08} Park SY, {\it et al.} (2008) DNA-programmable nanoparticle crystallization {\it Nature} {451}: 553-556.

\bibitem{gang08} Nykypanchuk D, Maye MM, van der Lelie D, Gang O. (2008) DNA-guided crystallization of colloidal nanoparticles. {\it Nature} {451}: 549-552. 

\bibitem{mclaughlin} Stewart KM, McLaughlin LW (2004) Four-arm oligonucelotide Ni(II)--cyclam-centered complexes as precursors for the generation of supramolecular assemblies.  {\it J. Am. Chem. Soc.} {126}: 2050-2057.


\bibitem{baglioni} {Tumpane} J, {\em et al.}  (2007)
    \newblock{Addressable high-information-density DNA nanostructures}
   {\em Chemical Physics Letters} {440}: {125-129}. 

\bibitem{seeman3}
Winfree E, Liu FR, Wenzler LA, Seeman NC (1998) Design and self-assembly of two-dimensional DNA crystals {\it Nature} {394}: 539-544.

\bibitem{crocker05} Biancaniello PL, Kim AJ, Crocker JC (2005) Colloidal interactions and self-assembly using DNA hybridization. {\em Phys. Rev. Lett.} {94}: 058302. 

\bibitem{kiang05} Harris NC, Kiang C-H (2005) Disorder in DNA-linked gold nanoparticle assemblies {\it Phys. Rev. Lett.} {95}: 046101.


\bibitem{zac05}
Zaccarelli E. {\it et al.} (2005) Model for reversible colloidal gelation {\it Phys. Rev. Lett.} {94}: 218301.   

\bibitem{pses92}  
Poole PH, Sciortino F, Essmann U, Stanley HE (1992) Phase behaviour of metastable water. {\it Nature} {360}: 324-328. 

\bibitem{ms98}
Mishima O, Stanley HE (1998) Decompression-induced melting of ice IV and the liquidÐliquid transition in water. {\it Nature} {392}: 164-168.


\bibitem{ms98review}
Mishima O, Stanley, HE (1998) The Relationship between Liquid, Supercooled and Glassy Water. {\em Nature} 396: 329-335.

\bibitem{debenedetti-review}
Debenedetti PG (2003) Supercooled and glassy water. {\em J. Phys.: Condens. Matter} 15: R1669ÐR1726.

\bibitem{brovchenko}
{Brovchenko} I, {Geiger} A, {Oleinikova} A (2003)
\newblock Multiple liquid-liquid transitions in supercooled water
{\em J. Chem. Phys.} {118}: 9473-9476. 
 

\bibitem{sa03}
Sastry S, Angell CA (2003) Liquid-liquid phase transition in supercooled silicon. {\it Nature Materials} {2}: 739-743.

\bibitem{saika} Saika-Voivod I, Sciortino F,  Grande T, Poole PH (2004) Phase diagram of silica from computer simulation.
{\it Phys. Rev. E} {70}: 061507.

\bibitem{polyamorphism}
Poole PH,  Grande T,  Angell CA, McMillan PF, (1997) Polymorphic Phase Transitions in Liquids and Glasses {\it Science} {17}: 322-323. 

\bibitem{fletcher}
Fletcher NH (1970) {\it The Chemical Physics of Ice.} (Cambridge University Press, Cambridge).

\bibitem{ss} Starr FW, Sciortino F (2006) Model for assembly and gelation of four-armed 
DNA dendrimers {\em J. Phys.: Condens. Matter} {18}: L347-L353.

\bibitem{lss} Largo J, Starr FW, Sciortino F (2007) Self-assembling DNA dendrimers: A numerical study. {\it Langmuir} {23}: 5896-5905.


\bibitem{lst}  Largo J, Tartaglia P, Sciortino F. (2007)  Effective nonadditive pair potential for lock-and-key interacting particles: The role of the limited valence. {\it Phys. Rev. E} {76}: 011402.


\bibitem{ising-CP} Tsypin MM, Blote HWJ (2000) Probability distribution of the order parameter for the three-dimensional Ising-model universality class: A high-precision Monte Carlo study. {\em Phys. Rev. E} {62}: 73-76.  

\bibitem{phosphorus}
Katayama Y, Mizutani T, Utsumi W, Shimomura O, Yamakata M, Funakoshi KA (1999) First-order liquid-liquid phase transition in phosphorus. {\it Nature} {403}: 170-173.

\bibitem{bs03}
Buldyrev SV,Stanley HE (2003) A system with multiple liquid-liquid
critical points. {\it Physica A} 330: 124-129.

\bibitem{kittel} Kittel C (2005) {\it Introduction to Solid State Physics.} (Wiley Publishers, Hoboken, NJ).

\bibitem{sgs91}
Sciortino F, Geiger A, Stanley HE (1991) Effect or defect on molecular mobility in liquid water.  {\em Nature} 354: 218-221.

\bibitem{sbs}
Starr FW, Bellissent-Funel M-C, Stanley HE.  (1999) Structure of
supercooled and glassy water.  {\it Physical Review E\/} 60: 1084-1087.

\bibitem{lg06}
Loerting T, Giovambattista N. (2006) Amorphous ices: experiments and numerical simulations. {\em J. Phys. Condens. Matt.} 18: R919.

\bibitem{franzese01}
Franzese G, Malescio G, Skibinsky A,  Buldyrev SV, Stanley, HE (2001) Generic Mechanism for Generating a Liquid-Liquid Phase Transition. {\em Nature} 409: 692-695.

\bibitem{xu05}
Xu L, Kumar P, Buldyrev SV, Chen S-H, Poole PH, Sciortino F, Stanley HE (2005) Relation between the Widom Line and the Dynamic Crossover in Systems with a Liquid-Liquid Critical Point. {\em Proc. Natl. Acad. Sci.} 102: 16558-16562.

\bibitem{xu06}
Xu L, Buldyrev SV, Angell CA, Stanley HE (2006) Thermodynamics and Dynamics of the Two-Scale Spherically Symmetric Jagla Ramp Model of Anomalous Liquids. {\em Phys. Rev. E} 74: 031108 (2006)

\bibitem{wilding}
Wilding NB (2001) Computer simulation of fluid phase transitions. {\it Am. J. Phys.} {69}: 1147-1155.


\end{thebibliography}
\end{document}